\documentclass[aps,pra,twocolumn,showpacs]{revtex4}
\usepackage{amsfonts,amssymb,amscd,amsmath}
\usepackage{graphicx}
\usepackage{epsfig}
\usepackage{latexsym}
\usepackage{amssymb}
\usepackage{subfigure}
\usepackage{color}

\begin{document}

\def\abs#1{ \left| #1 \right| }
\def\lg#1{ | #1 \rangle }
\def\rg#1{ \langle #1 | }
\def\lrg#1#2#3{ \langle #1 | #2 | #3 \rangle }
\def\lr#1#2{ \langle #1 | #2 \rangle }
\def\me#1{ \langle #1 \rangle }

\newcommand{\bra}[1]{\left\langle #1 \right\vert}
\newcommand{\ket}[1]{\left\vert #1 \right\rangle}
\newcommand{\bx}{\begin{matrix}}
\newcommand{\ex}{\end{matrix}}
\newcommand{\be}{\begin{eqnarray}}
\newcommand{\ee}{\end{eqnarray}}
\newcommand{\nn}{\nonumber \\}
\newcommand{\no}{\nonumber}
\newcommand{\de}{\delta}
\newcommand{\lt}{\left\{}
\newcommand{\rt}{\right\}}
\newcommand{\lx}{\left(}
\newcommand{\rx}{\right)}
\newcommand{\lz}{\left[}
\newcommand{\rz}{\right]}
\newcommand{\inx}{\int d^4 x}
\newcommand{\pu}{\partial_{\mu}}
\newcommand{\pv}{\partial_{\nu}}
\newcommand{\au}{A_{\mu}}
\newcommand{\av}{A_{\nu}}
\newcommand{\p}{\partial}
\newcommand{\ts}{\times}
\newcommand{\ld}{\lambda}
\newcommand{\al}{\alpha}
\newcommand{\bt}{\beta}
\newcommand{\ga}{\gamma}
\newcommand{\si}{\sigma}
\newcommand{\ep}{\varepsilon}
\newcommand{\vp}{\varphi}
\newcommand{\zt}{\mathrm}
\newcommand{\bb}{\mathbf}
\newcommand{\dg}{\dagger}
\newcommand{\og}{\omega}
\newcommand{\Ld}{\Lambda}
\newcommand{\m}{\mathcal}
\newcommand{\dm}{{(k)}}

\title{Quantum limits on detection sensitivity of a linear detector with feedback}
\author{Yang Gao}
\email{gaoyangchang@outlook.com} \address{Department of Physics,
Xinyang Normal University, Xinyang, Henan 464000, China}

\begin{abstract}
We show that the detection sensitivity of a linear detector is lower
bounded by some quantum limits. For the force detection, relevant
for atomic force microscopes, the lower bound is given by the
so-called ultimate quantum limit. For the displacement detection,
relevant for detecting gravitational waves, a generalized lower
bound is obtained.
\end{abstract}

\maketitle

\section{Introduction}

It is known that an optomechanical system, i.e., a mechanical
oscillator coupled to an optical cavity by radiation pressure, could
be a sensitive detector for very weak forces or tiny displacements.
Some examples are atomic force microscopes \cite{atom} and
interferometers for detecting gravitational waves \cite{gwave}. For
a typical optomechanical detector, the detection sensitivity is
lower bounded by the so-called standard quantum limit (SQL) due to
quantum noises \cite{qm,cave,clerk}.

However, the SQL itself is not a fundamental quantum limit.
Different schemes to overcome the SQL have been proposed, such as
frequency dependent squeezing (FDS) of the input beam
\cite{fds,uql}, cavity detuning (CD) \cite{cd}, variational
measurement (VM) \cite{vm}, quantum locking of the mirror
\cite{qlock}, coherent quantum noise cancelation (CQNC) \cite{cqnc},
and etc. It is shown in Ref. \cite{prove} that the fundamental
quantum limit for the force sensitivity of a linear detector is
given by the so-called ultimate quantum limit (UQL) in Refs.
\cite{uql,cd}. This UQL is related to the dissipation mechanism of
the oscillator, via the absolute value of the imaginary part of the
inverse mechanical susceptibility. It also holds for the cases with
coherent quantum control and/or with simple direct quantum feedback
\cite{cqc,qfb}. Now we try to extend the result in Ref. \cite{prove}
to the cases with more complicated quantum feedback loop, such as
quantum locking of the mirror via a control cavity \cite{qlock}.

On the other hand, for the displacement sensitivity, we find that
the usual UQL, given by the absolute value of the imaginary part of
the mechanical susceptibility, can be overcome by some devised
schemes. Moreover, a generalized quantum limit on the displacement
sensitivity can be obtained following the similar arguments for
quantum limit on the force sensitivity.

The paper is organized as follows. In Section II, we prove the UQL
for the force sensitivity of a linear detector in the presence of
direct quantum feedback by using the general linear-response theory.
The results are then adjusted to establish a general quantum limit
on the displacement sensitivity. Section III contains some
illustrating examples for the above results. At last, a short
summary is given in Section IV.

\begin{figure}[t!]
\centering \subfigure[]{
\includegraphics[width=0.4\columnwidth,angle=-90]{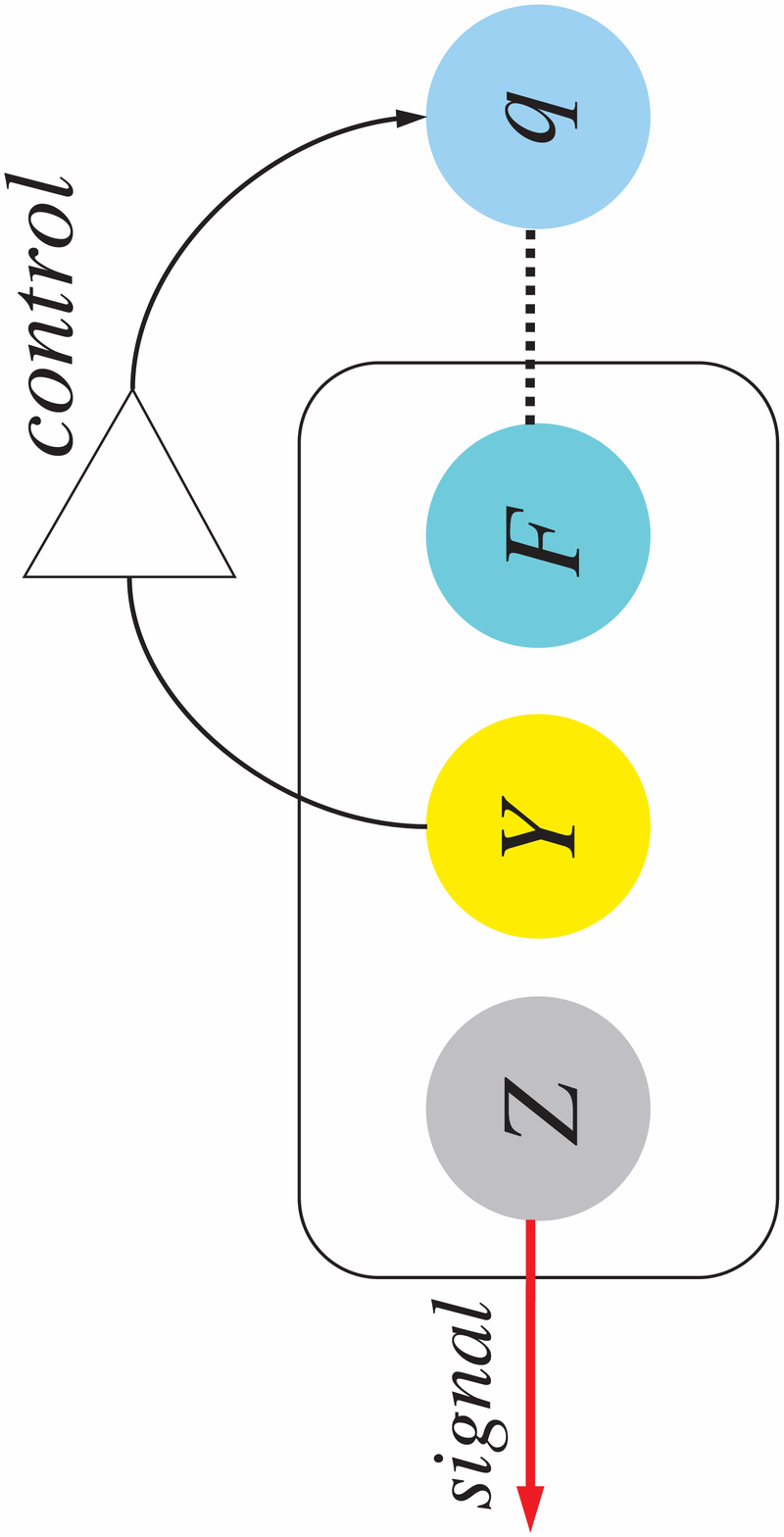}}\hspace{0.01in}
\subfigure[]{
\includegraphics[width=0.4\columnwidth,angle=-90]{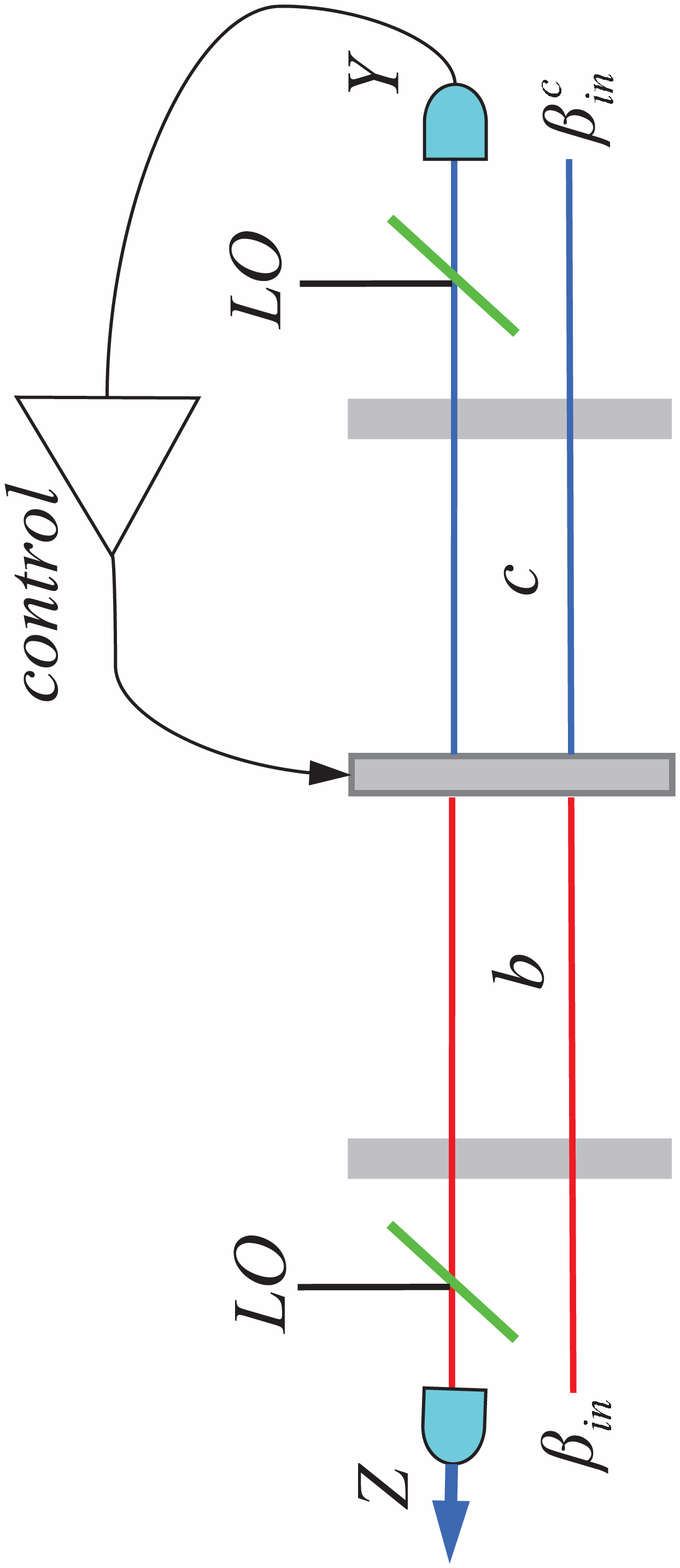}}\hspace{0.01in}
\caption{Schematics of (a) a generic linear-response detector and
(b) an optomechanical detector. The outputs $Y,Z$ from the
photodiodes are modulated by the local oscillator (LO) phases. }
\end{figure}

\section{Detection sensitivity of a linear detector with feedback}

We consider a generic linear-response detector (see Fig. 1(a)). It
is described by some unspecified Hamiltonian $H_d$, and has both an
input operator, represented by an operator $F$, and an output
operator, represented by an operator $\m Z$. The detector is used to
estimate a classical force $f(t)$ applied on a mechanical
oscillator, which is represented by the Hamiltonian $H_m$. The input
operator $F$ is coupled with the mechanical oscillator via the
interaction Hamiltonian, $H_{int}=- q(f+g F)$, where $q$ is the
displacement operator of the oscillator. The output operator $\m Z$
(e.g., the output optical quadrature) is related to the readout
quantity at the output of the detector (e.g., the output current of
the photodiode), from which the information of the classical force
can be inferred.

The total Hamiltonian is given by $H=H_m+H_d+H_{int}$. Treating
$H_{int}$ as the perturbation, an arbitrary operator $O$ in the
Heisenberg picture is obtained by \be O(t)= \mathcal U^\dg O_0(t)
\mathcal U, \quad \mathcal U(t)=\mathcal T e^{-i\int_{-\infty}^t
H^0_{int}(t) dt}, \label{his} \ee where $O_0(t)$ denotes the
operator in the interaction picture, and the symbol $\mathcal T$
means the time-ordered product. For the linear operators $q$, $F$,
and $\m Z$ (with the c-number commutators), Eq. (\ref{his}) gives
the equation of motion in the frequency domain \cite{qlock}, \be q
&=& q_0 + q_f + g \chi_{qq} F, \nn F &=& F_0 + g \chi_{F\!\!\!\ F}
q, \nn Z &=& Z_0 + g q, \label{lre} \ee where $q_f=\chi_{qq} f$, and
$Z$ is the rescaled output operator via $\m Z(\og)=\chi_{\m Z\!\!\!\
F}(\og) Z(\og) $. The susceptibility $\chi_{O_1\!\!\!\ O_2}$ is
defined by the c-number commutator, $ \chi_{O_1 \!\!\!\ O_2}(t)=i
\theta(t)[O_1(t),O_2(0)]$.

Solving the first two equations and substituting into the third one
of Eq. (\ref{lre}), we have the output operator in terms of the
unperturbed operators, \be Z=Z_0+g{q_0+q_f+ g\chi_{qq} F_0 \over 1 -
g^2\chi_{qq}\chi_{F\!\!\!\ F}}. \ee It then gives an unbias
estimator $q_\zt{est}$ of $q_f$, $q_\zt{est}=q_f+\hat q_f$, where
the added noise $\hat q_f$ is \be \hat{q}_f=q_0 + g \chi_{qq} F_0 +
\frac{Z_0}{g}(1-g^2 \chi_{qq} \chi_{F\!\!\!\ F}). \ee Here the first
term is the intrinsic mechanical noise, the second term is the
backaction noise from the oscillator, and the third term is the shot
noise at the output. Neglecting the intrinsic mechanical noise from
now on, the force sensitivity can be characterized by the power
spectrum $S_{\hat{q}_f\hat{q}_f}$ defined by the correlation \be
\frac{1}{2}\me{O_1(\og)O_2^\dg(\og')+O_2^\dg(\og')O_1(\og)} =
S_{O_1O_2} \delta(\og-\og'). \ee Using the uncertainty relation for
$F_0$ and $Z_0$, the optimization of $S_{\hat{q}_f\hat{q}_f}$ over
the coupling strength $g$ gives \be S_{\hat{q}_f\hat{q}_f}\ge
|\chi_{qq}^I|, \ee where $\chi_{qq}^I$ is the imaginary part of the
susceptibility $\chi_{qq}=\chi_{qq}^R+i\chi_{qq}^I$. Equivalently,
the force sensitivity in terms of $f$ yields \be S_f  =
\frac{S_{\hat{q}_f\hat{q}_f}}{|\chi_{qq}|^2} \ge
|\bar\chi_{qq}^I|,\ee where $\bar \chi_{qq}=1/\chi_{qq}$ is the
inverse mechanical susceptibility. This is the UQL for the force
sensitivity.

The above result has incorporated the effect of coherent quantum
control \cite{cqc}, such as the CNQC scheme. As for the direct
quantum feedback control \cite{qfb}, a control signal $\m Y(t)$ is
fed back to the system, see Fig. 1(a). For a generic operator $O$,
it introduces an additional term to the equation of motion, \be \dot
O_{ fb}(t) = i \int_{-\infty}^t dt'\widetilde{\ld}(t-t') \m Y(t')
[\m P(t),O(t)], \ee where $\widetilde{\ld}(t)$ is the feedback
transfer function, $\m P(t)$ is the control operator. If the linear
control operator $\m P$ is of the mechanical oscillator, $[\m P,
q]=const.$ and $[\m P, F]=0$. In the frequency domain, we have
$q_{fb} = \ld Y$, where $\ld$ is the rescaled transfer function.

It has been shown in Ref. \cite{prove} that if the measured signal
is the same as the control signal fed back to the system ($\m Z=\m
Y$), the force sensitivity does not change. On the other hand, if
the measured signal is different from the control signal ($\m Z \neq
\m Y$), such as the cases of quantum locking of the mirror, the
equation of motion becomes \be q &=& q_0 + q_f + g \chi_{qq} F + \ld
Y, \nn F &=& F_0 + g \chi_{FF} q, \nn Y &=& Y_0 + g q,\nn Z &=&
Z_0+g q.  \ee We find that \be q = { q_0 + q_f + \ld Y_0+g\chi_{qq}
F_0 \over 1 - g \ld -g^2\chi_{qq} \chi_{FF} }. \ee The estimator of
$q_f$ deduced from the measured signal $Z$ thus is
$q_\zt{est}=q_f+\hat q_f$ with the added noise \be \hat{q}_f=q_0 + g
\chi_{qq} F_0+ \ld Y_0 + \frac{Z_0}{g}(1-g\ld-g^2 \chi_{qq}
\chi_{FF}). \ee We find that for $Z=Y$, the above equation reduces
to Eq. ().

To find a lower bound to the power spectrum of $\hat{q}_f$, we first
optimize $S_{\hat q_f \hat q_f}$ over the transfer function $\ld$
and the coupling strength $g$, respectively. This leads to \be
S_{\hat q_f \hat q_f} \ge {2 (A \chi_{qq}^R + B \chi_{qq}^I +
|\chi_{qq}| \sqrt{C}) \over S_{YY}+S_{ZZ}}, \label{abc} \ee where
the notations are \be A &=& S_{ZF}^R S_{YY} + S_{YF}^R
S_{ZZ}-\chi_{FF}^R S_{YY}S_{ZZ} \nn B&=&S_{ZF}^I S_{YY} + S_{YF}^I
S_{ZZ} + \chi_{FF}^I S_{YY}S_{ZZ} \nn C&=& [ S_{FF} (S_{YY} +
S_{ZZ})+ |\chi_{FF}|^2 S_{YY} S_{ZZ} \nn && - (S_{YF}^I -S_{ZF}^I)^2
- (S_{YF}^R -S_{ZF}^R)^2 \nn && + 2S_{YY}( \chi_{FF}^I S_{ZF}^I  -
\chi_{FF}^R S_{ZF}^R) \nn && + 2 S_{ZZ} ( \chi_{FF}^I S_{YF}^I -
\chi_{FF}^R S_{YF}^R)] S_{YY}S_{ZZ}. \label{cc} \ee Next, we note
that $ [Y_0(t),Y_0(t')]=[Z_0(t), Z_0(t')]=0 $ at all times, in order
for $Y_0(t)$ and $Z_0(t)$ to represent experimental data strings. It
immediately implies that $\chi_{Y\!\!\!\ Y}=\chi_{Z\!\!\!\ Z}=0$.
Also, the causality principle imposes that the outputs $Y_0(t)$ and
$Z_0(t)$ should not depend on the input $F_0(t')$ for $t<t'$, and
therefore $\chi_{F\!\!\!\ Y}(\og)=\chi_{F\!\!\!\ Z}(\og)=0$.
Furthermore, $Y_0$, $Z_0$, and $F_0$ should satisfy the uncertainty
relation \cite{qm,prove} implied by the positivity of the matrix
$M_{jk}=S_{jk} \pm i[\chi_{jk}-\chi^*_{kj}]/2$ with the indexes
$j,k=Y_0,Z_0,F_0$, \be
M=\lz \bx S_{YY} & 0 & S_{YF}\pm {i / 2} \\ 0 & S_{ZZ} & S_{ZF} \pm  {i / 2} \\
S_{YF}^*\mp  {i / 2} & S_{ZF}^*\mp  {i / 2} & S_{FF} \mp \chi_{FF}^I
\ex \rz \ge 0 \ee or equivalently, \be && S_{FF} S_{YY} S_{ZZ} \ge
{S_{YY}+ S_{ZZ} \over 4}\pm (S_{ZF}^I S_{YY}+ S_{YF}^I S_{ZZ}) \nn
&& \indent + |S_{ZF}|^2 S_{YY} +|S_{YF}|^2 S_{ZZ} \pm \chi_{FF}^I
S_{YY} S_{ZZ}. \ee Here the relations
$\chi_{YF}-\chi^*_{FY}=\chi_{ZF}-\chi^*_{FZ}=1$ and
$S_{jk}=S^*_{kj}$ have been used.

Substituting this inequality into Eq. (\ref{cc}), by some
calculations we find \be C \ge A^2+\lx B \pm
\frac{S_{YY}+S_{ZZ}}{2}\rx^2 \nn = A^2+\lx |B| +
\frac{S_{YY}+S_{ZZ}}{2}\rx^2. \ee Then from Eq. (\ref{abc}), we
finally obtain \be S_{\hat q_f \hat q_f} \ge |\chi_{qq}^I|, \ee
where the inequalities $a_1 x_1 +\sqrt{(a_1^2+a_2^2)(x_1^2+x_2^2)}
\ge |a_2 x_2|$ and $|a| \ge a$ have been used. The force sensitivity
$S_f$ is then given by \be S_f \ge |\bar \chi_{qq}^I|, \ee which is
the main result of this paper. Similar result can be obtained if the
control operator $\m P$ is from the detector. The UQL is thus
established in the presence of direct quantum feedback.

Now we show whether the displacement sensitivity $S_q$ is lower
bounded by $|\chi_{qq}^I|$, namely, the usual UQL for the
displacement sensitivity. It is known that this UQL is valid in the
weak coupling limit and/or for a detector with a large power gain
\cite{qm,clerk}. In general, the displacement sensitivity is given
by the power spectrum of the estimator $q_\zt{est}$ of $q$ deduced
from the measured signal $Z$, $q_\zt{est}={Z / g}=q+\hat q$, where
the added noise is \be \hat q = \frac{\hat q_f}{1-g \ld-
g^2\chi_{qq}\chi_{FF}}.\ee Then we get \be S_q &=& S_{\hat q\hat q}
= \frac{S_{\hat q_f \hat q_f}}{|1-g \ld- g^2\chi_{qq}\chi_{FF}|^2}
\nn &\ge & \frac{|\chi_{qq}^I|}{|1-g \ld- g^2\chi_{qq}\chi_{FF}|^2},
\label{disp} \ee which can therefore overcome the UQL by properly
adjusting the detector parameters to make the above denominator
larger than one, as illustrated in the subsequent examples. We also
note that in the weak coupling limit, the denominator is
approximately equal to one, and Eq. (\ref{disp}) reduces to the UQL.

\section{Cavity detuning and quantum locking of the mirror}

To exemplify the results in the last section, we take the
optomechanical detector as an example. The optomechanical detector
consists of a high quality Fabry-Perot cavity, with a fixed
transmissive mirror in front of the cavity, and a moveable,
perfectly reflecting mirror $m$ at the back (see Fig. 1(b)). The
cavity field described by the annihilation operator $b=(b_1+i
b_2)/\sqrt{2}$ with resonant frequency $\og_b$ is fed with a driving
laser $\bt_{in}$. The aim is to estimate a classical force $f(t)$
acting on the moveable mirror described by the annihilation operator
$a=(q+ip)/\sqrt 2$ with frequency $\Omega$. In the rotating frame at
the frequency $\og_0$ of the driving laser, the system is described
by the Hamiltonian, \be H &=& H_m+H_d+H_{int} \nn &=& \Omega a^\dg
a+[\Delta b^\dg b+i\sqrt \ga (\bt_{in} b^\dg-\bt_{in}^* b)] \nn &&
-q[f+g_{om}(b^\dg b-\me{b^\dg b})], \ee where $\Delta=\og_b-\og_0$
is the cavity detuning, and $g_{om}$ is the optomechanical coupling
strength. Taking into account the thermal noises, the equations of
motion are given by the quantum Langevin equations \cite{qop}, \be
\dot a &=& -i\Omega a-\frac{\Gamma}{2}a+[f+g_{om}(b^\dg b-\me {b^\dg
b})]/\sqrt{2}+\sqrt{\Gamma}a_{in}, \nn \dot b &=& -i\Delta b
-\frac{\ga}{2} b+i g_{om} q b+\sqrt{\ga}(\bt_{in}+b_{in}),
\label{lin} \ee where $\Gamma (\ga)$ and $a_{in} (b_{in})$ are the
decay rate and thermal noise operator for the oscillator (cavity),
respectively. The noise correlations are given by $\me
{a_{in}(t)a_{in}^\dg(t')}=(n_{th}+1) \de (t-t')$ and $\me
{b_{in}(t)b_{in}^\dg(t')}= \de (t-t')$, where $n_{th}$ is the
thermal occupancy of the mechanical reservoir.

Under the condition of strong laser driving, we can linearized Eq.
(\ref{lin}) around the steady state, $\me a=0$ and $\me
b=\bt=\sqrt{\ga}\bt_{in}/(\ga/2+i\Delta)$, by splitting $a \to \me a
+ a$ and $b \to \me b + b$. Neglecting the nonlinear terms, we have
\be \dot {\bb x} = {\bb A \bb x}+\bb w, \label{eom} \ee where the
variables $\bb x=(q,p,b_1,b_2)^\top$, the input $\bb
w=(\sqrt{\Gamma}q_{in},f+\sqrt{\Gamma}p_{in},
\sqrt{\gamma}b_1^{in},\sqrt{\gamma}b_2^{in})^\top$, and the
matrix \be \bb A= \lx \bx -{\Gamma \over 2} & \Omega & 0 & 0 \\
-\Omega & -{\Gamma \over 2} & g & 0 \\ 0 & 0 & -{\gamma \over 2} &
\Delta \\ g & 0 & -\Delta & -{\gamma \over 2} \ex \rx \ee in terms
of the effective optomechanical coupling strength $g=\sqrt 2 g_{om}
\bt $. The stability of this linearized system is guaranteed by the
requirement that the real part of all the eigenvalues of $\bb A$
must be nonpositive. The classical force is then estimated from the
output current $I_{out}$ of a photodiode that is linearly
proportional to a certain optical quadrature of the output field, $
I_{out} \propto \m Z = b_1^{out}\sin \phi +b_2^{out}\cos\phi$, where
the output field is obtained by the input-output relation
$b_{out}=\sqrt{\ga}b-b_{in}$, and $\phi$ is the adjustable readout
quadrature angle via the local oscillator phase. For the stationary
state, Eq. (\ref{eom}) can be simply solved in the frequency domain.
The corresponding numerical results for the detection sensitivities
are shown in Fig. 2. It can be seen that for a detuned cavity, the
force sensitivity is lower bounded by the relevant UQL, $S_f \ge
|\bar\chi_{qq}^I|= \og\Gamma/\Omega $ with
$\chi_{qq}={\Omega}\lz{(\Gamma/2-i\og)^2+\Omega^2}\rz^{-1}$, and the
displacement sensitivity can overcome the corresponding UQL, namely,
$S_q < |\chi_{qq}^I|$.

For the displacement sensitivity, it is also found that the usual
UQL is still valid for a resonant cavity. Because the susceptibility
$\chi_{FF} \propto \Delta$ with the operator $F=b_1$ for the
optomechanical detector, and the denominator in Eq. (\ref{disp})
becomes unity for $\Delta=0$. However, there is another possibility
to overcome the usual UQL for a resonant cavity with the help of
quantum feedback ($\ld \neq 0$), such as quantum locking of the
mirror. In this scheme, the addition of a feedback loop is used to
suppress radiation pressure effects by freezing the motion of
mirror. The mirror motion is monitored by a feedback force $\propto
\m Y = c_1^{out}\sin \theta+c_2^{out}\cos \theta$ from another
resonant cavity made of the movable mirror $m$ and a fixed
transmissive, reference mirror. The control system is governed by
the Hamiltonian, \be H_{c}=i\sqrt \ga (\bt_{in}^c c^\dg-
\bt_{in}^{c\!\
*} c)+g_{om}q(c^\dg c-\me{c^\dg c}), \ee
where the control cavity field $c$ with frequency $\og_c$ is in
resonance with the driving laser $\bt_{in}^c$, and takes the same
decay rate $\ga$ as the main cavity.

\begin{figure}[t!]
\centering \subfigure[]{
\includegraphics[width=0.48\columnwidth]{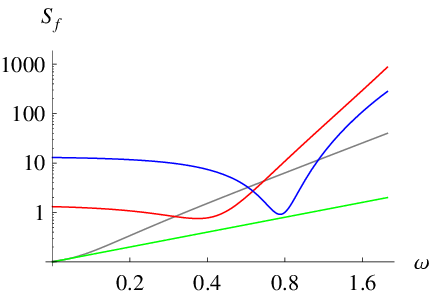}}\hspace{0.01in}
\subfigure[]{
\includegraphics[width=0.48
 \columnwidth]{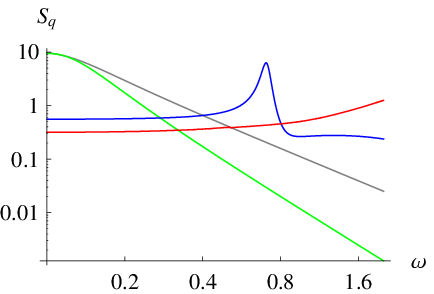}}\hspace{0.01in}
\caption{Plots of the force sensitivity (a) and the displacement
sensitivity (b) with respect to the detection frequency for the
optomechanical detector. Grey: SQL; green: UQL; blue: detuned cavity
for $\ga=2$, $g=5$, $\Delta=-5$, and $\tan \phi=-1$; red: resonant
cavity with feedback for $g=\zt g/2=1$, $\phi=0$, (a) $\ga=5$, $\tan
\theta=2$, and (b) $\ga=2$, $\theta=0$. Here the detection
sensitivities are optimized over the feedback transfer function
$\ld(\og)$. The common parameters are $\Omega=\Gamma=0.1$.}
\end{figure}

The linearized equations of motion take the same form of Eq.
(\ref{eom}), where \be \bb x&=&(q,p,b_1,b_2,c_1,c_2)^\top, \nn \bb
A &=& \lx \bx -{\Gamma \over 2} & \Omega & 0 & 0 & 0 & 0\\
-\Omega & -{\Gamma \over 2} & g & 0 & -\zt g_1 & \zt g_2
\\ 0 & 0 & -{\gamma \over 2} & 0&0 &0
\\ g & 0 & 0 & -{\gamma \over 2} &0&0 \\ 0& 0& 0& 0& -{\gamma \over 2} &0
\\ -\zt g & 0& 0& 0& 0 & -{\gamma
\over 2} \ex \rx, \\ \bb w&=&(\sqrt{\Gamma}q_{in},f_\zt{eff},
\sqrt{\gamma}b_1^{in},\sqrt{\gamma}b_2^{in},\sqrt{\gamma}c_1^{in},\sqrt{\gamma}c_2^{in})^\top
\nonumber \ee in terms of the effective force
$f_\zt{eff}=f+\sqrt{\Gamma}p_{in}-\ld(c_1^{in}\sin
\theta+c_2^{in}\cos \theta)$, the effective coupling strengthes
$g=2g_{om}\bt_{in}\sqrt{2/\ga}$, $\zt
g=2g_{om}\bt_{in}^c\sqrt{2/\ga}$, $\zt g_1=\zt g- \ld\sqrt{\ga}\sin
\theta$, and $\zt g_2=\ld\sqrt{\ga} \cos \theta$. The detection
sensitivities obtained from the output current $ I_{out} \propto \m
Z = b_1^{out}\sin \phi +b_2^{out}\cos\phi$ are also plotted in Fig.
2. It shows that for a resonant cavity with a feedback loop, the
force sensitivity is still lower bounded by the UQL, and the
displacement sensitivity can overcome the ususal UQL by properly
choosing detector parameters.

\section{Conclusion}

We have proven the UQL for the force sensitivity of a generic
linear-response detector in the presence of direct quantum feedback
by using the general linear-response theory. We have shown that the
usual UQL for the displacement sensitivity can be overcome by some
devised schemes. By adopting the arguments for the force
sensitivity, a generalized quantum limit is obtained for the
displacement sensitivity. To illustrate the utilities of these
quantum limits, we have taken a detuned cavity and a resonant cavity
with a feedback loop as two specific examples. We believe that our
results show the ways to improve the performance of high-sensitivity
detection schemes.

YG acknowledges the support from NSFC Grant No. 11304265.


\end{document}